\documentclass[12pt]{article}
\usepackage{amsmath,amsfonts, amssymb, braket, bbold}
\usepackage{tikz}
\usetikzlibrary{trees,er,snakes,shapes,mindmap}
\usepackage{titlesec}
\titleformat{\section}
 {\normalfont\fontfamily{put}\fontsize{12pt}{16pt}\bfseries\color{black}}
{\thesection}{1em}{}
\def \beq  {\begin{equation}}
\def \eeq  {\end{equation}}
\def \beqar {\begin{eqnarray}}
\def \eeqar {\end{eqnarray}}
\allowdisplaybreaks
\def\sqr#1#2{{\vcenter{\vbox{\hrule height.#2pt
\hbox{\vrule width.#2pt height#1pt \kern#1pt
\vrule width.#2pt}\hrule height.#2pt}}}}

\def\S {{\cal S}}

\def\vx {{\vec x}}

\def\Tr {{\rm Tr}}

\def\vx {{\vec x}}

\def\del {\partial}

\def\e {\epsilon}

\def\A {{\cal A}}

\def\C {{\cal C}}

\def\E {{\cal E}}

\def\H {{\cal H}}

\def\M{{\cal M}}

\def\half{\textstyle{1\over 2}}

\mathchardef\mhyphen="2D
\begin{document}
%%%%%%%%%%%%%%%%%%%%%%%%%%%%%%%%%%%%%%%%%%%%%%%
%%%%%%%%%%%%%%%%%%%%%%%%%%%%%%%%%%%%%%%%%%%%%%%
\fontfamily{bch}\fontsize{12pt}{16pt}\selectfont
%\fontfamily{pnb}\fontsize{12pt}{16pt}\selectfont
%\fontfamily{pzc}\fontsize{14pt}{16pt}\selectfont
%\fontfamily{pbk}\fontsize{12pt}{16pt}\selectfont
%\fontfamily{cmr}\fontsize{11pt}{15pt}\selectfont
%\fontfamily{put}\fontsize{11pt}{16pt}\selectfont
%\fontfamily{lmss}\fontsize{11pt}{16pt}\selectfont
%\fontfamily{phv}\fontshape{ro}\fontsize{11pt}{14pt}\selectfont
%\fontfamily{ptm}\fontseries{m}\fontshape{r}\fontsize{12pt}{16pt}\selectfont
%\fontfamily{pnc}\fontseries{m}\fontshape{r}\fontsize{11pt}{15pt}\selectfont
%\fontfamily{ppl}\fontseries{m}\fontshape{r}\fontsize{11pt}{15pt}\selectfont
%\usefont{T1}{phv}{m}{it}
%%%%%%%%%%%%%%%%%%%%%%%%%%%%%%%%%%%%%%%%%%%%%%%
%%%%%%%%%%%%%%%%%%%%%%%%%%%%%%%%%%%%%%%%%%%%%%%
\def \CMP {{Commun. Math. Phys.}}
\def \PRL {{Phys. Rev. Lett.}}
\def \PL {{Phys. Lett.}}
\def \NPBProc {{Nucl. Phys. B (Proc. Suppl.)}}
\def \NP {{Nucl. Phys.}}
\def \RMP {{Rev. Mod. Phys.}}
\def \JGP {{J. Geom. Phys.}}
\def \CQG {{Class. Quant. Grav.}}
\def \MPL {{Mod. Phys. Lett.}}
\def \IJMP {{ Int. J. Mod. Phys.}}
\def \JHEP {{JHEP}}
\def \PR {{Phys. Rev.}}
\def \JMP {{J. Math. Phys.}}
\def \GRG{{Gen. Rel. Grav.}}
%%%%%%%%%%%%%%%%%%%%%%%%%%%%%%%%%%%%%%%%%%%%%%%
%%%%%%%%%%%%%%%%%%%%%%%%%%%%%%%%%%%%%%%%%%%%%%%
\begin{titlepage}
\null\vspace{-62pt} \pagestyle{empty}
\begin{center}
%\rightline{CCNY-HEP-18/4}
%\rightline{August 2018}
\vspace{1.3truein} {\large\bfseries
Eductions of Edge Mode Effects}\\
%\vskip .15in
%{\Large\bfseries ~ }\\
\vskip .5in
{\Large\bfseries ~}\\
%%%%%%%%%%%%%%%%%%%%%%%%%%%%%%%%%%%%%%%%%%%%%%%
%%%%%%%%%%%%%%%%%%%%%%%%%%%%%%%%%%%%%%%%%%%%%%%
{\large\sc V.P. Nair}\\
\vskip .2in
{\sl Physics Department,
City College of the CUNY\\
New York, NY 10031}\\
 \vskip .1in
\begin{tabular}{r l}
{\sl E-mail}:&\!\!\!{\fontfamily{cmtt}\fontsize{11pt}{15pt}\selectfont vpnair@ccny.cuny.edu}\\
\end{tabular}
\vskip 1.5in
%%%%%%%%%%%%%%%%%%%%%%%%%%%%%%%%%%%%%%%%%%%%%%%
%%%%%%%%%%%%%%%%%%%%%%%%%%%%%%%%%%%%%%%%%%%%%%%
\centerline{\large\bf Abstract}
\end{center}
Edge modes in gauge theories, whose {\it raison d'\^etre} is 
in the nature of the test functions used for imposing the Gauss law, have implications in many physical contexts. I discuss two such cases: 1) how edge modes are related to
the interface term in the BFK formula and how they generate the so-called contact term for entanglement entropy in gauge theories, 2) how they describe the dynamics of particles in generalizing the Einstein-Infeld-Hoffmann approach
to particle dynamics in theories of gravity.
\vskip .15in
\noindent
To be published in {\it Particles, Fields and Topology: Celebrating
A.P. Balachandran}, a Festschrift volume for A.P. Balachandran
(World Scientific Publishing Co., Singapore).
\end{titlepage}
%%%%%%%%%%%%%%%%%%%%%%%%%%%%%%%%%%%%%%%%%%%%%%%
%%%%%%%%%%%%%%%%%%%%%%%%%%%%%%%%%%%%%%%%%%%%%%%
%\fontfamily{put}\fontsize{12pt}{17pt}\selectfont
\pagestyle{plain} \setcounter{page}{2}
\section{Introduction}
For many students of high energy physics in the early 1980s in Syracuse, afternoons repeated a familiar pattern. Balachandran, or Bal as everyone referred to him, would appear 
at the door of the tiny office I shared with a fellow graduate student,
in his signature attire, the greenish-grey sweater with
elbow patches, coffee cup in hand, saying ``Let us discuss".
We would walk down the hallway, collecting more students, 
sometimes visitors, to gather in Room 316, made famous by Fedele Lizzi, in his contribution to this volume. All matters great and small would come
up for discussion. 
And some days we would then repair to his home where Indra's
tolerance and the excellent food would let us talk late into the evening.
Bal was always fascinated by the mathematical side of
things in physics. He, and his students and collaborators, had just been
working on applying the coadjoint orbit actions to various problems
when I arrived in Syracuse. This was also a time of
an effervescence of many ideas in particle physics:
the role of topology in physics was beginning to be appreciated, 
monopoles, solitons, instantons made frequent appearances in papers,
anomalies were still intriguing, Bal talked of why he felt the
effective action for anomalies constructed by Wess and Zumino should be
important (prescient comments as it turned out) and the long dormant idea of Skyrmions was just about to be revived, 
Over time, all of us, his students and collaborators, have branched out in different ways, but a fascination and engagement with the mathematical 
side of things, from the Syracuse days with Bal,
have remained the leitmotif of our work.

In appreciation, in this article dedicated to Bal's 85th birthday,
I would like to highlight a couple of effects related to one of his 
many favorite topics, namely, edge modes in gauge theories
and gravity.

\section{Entanglement in a gauge theory and the `contact term'}

As is well known, in a gauge theory, physical states are annihilated by
the Gauss law operator $G(\theta )$ defined with test functions $\theta (x)$ which vanish at the boundary of the space under consideration. 
In this case $e^{i G (\theta )}$ generates gauge transformations
$e^{i \theta }$ which become the identity on the boundary.
The same operator $e^{i G({\tilde \theta})}$, with ${\tilde \theta}$ 
which do not vanish on the boundary, generate physical states;
these are the edge modes. In the 1990s, Bal and collaborators 
explored the properties of such states for the Maxwell-Chern-Simons
theory, as well as in more general contexts \cite{{gen}, {APB}}.
More recently, as interest in entanglement has increased,
it has become clear that edge modes do make a contribution, the so-called contact term, to the entanglement entropy \cite{AKN}.
This is the facet of edge modes that I would first like to highlight.

We will consider the Maxwell theory in 2+1 dimensions as this suffices to illustrate the main point. In a Hamiltonian framework, we can take
$A_0 =0$. The spatial components of the gauge potential $A_i$
and the electric field $E_i$, ($i = 1,2$) can be parametrized as
\beq
A_i = \del_i \theta + \epsilon_{ij} \del_j \phi, \hskip .3in
E_i = {\dot A_i} = \del_i {\sigma} + \epsilon_{ij} \del_j {\Pi}
\label{edge1}
\eeq  
We consider the spatial manifold $\M$ to be a square which is
separated, by a straight line interface, into two rectangular regions which we will label as ${\rm L}$ and ${\rm R}$. The idea is to split the fields into three terms each, a field
in ${\rm L}$ which vanishes on the interface, a field in ${\rm R}$ which
also vanishes on the interface, and a field on the interface itself.
Focusing on the region
${\rm L}$ first, the fields can be split as
\beq
\chi_{\rm L} (x) = {\tilde \chi}_{\rm L}(x) + \int_{\del {\rm L}} \chi_{0} (y)\,n \cdot \del_y G_{\rm L} (y,x)
\label{edge2}
\eeq
where $\chi =$ $\theta$, $\phi$, $\sigma$, $\Pi$.
${\tilde \chi}_L$ vanishes on the boundary of $\M$ as well
as on the interface
between ${\rm L}$ and ${\rm R}$. $\chi_{0} $ denotes the value
of $\chi$ on the interface and $G_{\rm L} (y, x)$ is the Green's function for the
Laplacian, again obeying Dirichlet (vanishing) conditions on
the boundary of $\M$ and on the interface $\del {\rm L}$.
In (\ref{edge2}), the value of the field on the interface, namely $\chi_0$,
is continued into the interior of ${\rm L}$ by 
Laplace's equation and Green's theorem, so that
\beq
\nabla^2_x \, \int_{\del {\rm L}} \chi_{0} (y)\,
n \cdot G_{\rm L} (y,x) = 0
\label{edge3}
\eeq
This does not introduce any functional degrees of freedom in addition to $\chi_0$,
the freedom of choosing arbitrary values of the field in ${\rm L}$ is
contained in ${\tilde \chi}_{\rm L}$. Equation (\ref{edge2}) gives a general
parametrization of the fields in ${\rm L}$.

Consider now the phase space path integral where, for the constraint
$\nabla \cdot E \approx 0$, we choose the conjugate
constraint $\nabla \cdot A \approx 0$ (Coulomb gauge). The path integral is given by
\beqar
Z &=& \int d\mu~ \delta (\nabla \cdot E)~\delta(\nabla \cdot A)~ {\det (-\nabla^2)} ~e^{i \S}\nonumber\\
\S&=& \int d^3x \left[ E_i {\dot A_i} - \H \right]
\label{edge4}
\eeqar
where $\H$ is the Hamiltonian density.
The canonical two-form $\int \delta E_i \, \delta A_i$ serves to define the
phase space measure $d\mu$ in (\ref{edge4}). The canonical one-form
$\A= 
\int E_i \, \delta A_i$ is given in terms of the parametrization
(\ref{edge1}) as
\beq
\A = \int_{\rm L} \left[ (-\nabla^2 {\tilde \sigma}_{\rm L})\, \delta {\tilde \theta}_{\rm L}
+ {\tilde \Pi}_{\rm L} \, \delta B_{\rm L} \right]
+ \int_{\del {\rm L}} \E_{0}\, \delta \theta_{0} (x) 
+ \int_{\del {\rm L}}  Q_{0}\, \delta \phi_{0} (x) 
\eeq
where $B = -\nabla^2 \phi$ is the magnetic field and
\beqar
\E_{0}(x)&=& \int_y \sigma_{0} (y) M_{\rm L}(y,x)   + {\del_{\tau}} \Pi_{0}(x) \nonumber\\
Q_{0}(x) &=& \int_y \Pi_0 (y) M_{\rm L} (y,x)  - {\del_{\tau}} \sigma_{0} (x)
\label{edge6}
\eeqar
Here $\del_\tau$ signifies the tangential derivative
and
$M_{\rm L}(x, y) = n\cdot \del_x \,n\cdot \del_y G_{\rm L} (x,y)$
(with $x, y$ on $\del {\rm L}$)  is the Dirichlet-to-Neumann operator for the geometry we have. It is easily verified that this is essentially
$(\sqrt{-\nabla^2})_{x, y}$. As a result, we can see that
$\E_0$ and $Q_0$ are related by
$\C = \del_y \int \E_{{\rm I}}(x)\, M^{-1}(x,y) + Q_ {{\rm I}}(y)= 0$.
This is another constraint in the problem, it is
due to the freedom in defining $B$.
Notice that $B = -\nabla^2 (\phi + f ) = -\nabla^2 \phi$, if $\nabla^2 f =0$.
Nontrivial choices of such functions exist, for example, as
\beq
f (x) = \int_{\del {\rm L}}  f_0 (y) \, n\cdot \del_y G_{\rm L}  (y, x)
\label{edge7}
\eeq
The constraint $\C$ encodes this additional ``gauge freedom", which is really
an ambiguity of the parametrization (\ref{edge1}).
We can use the freedom of $f$ to set $\phi_0$ to zero, i.e., choose
$\phi_0 \approx 0$ as the constraint conjugate to $\C$.
Eliminating $\C$ and $\phi_0$ by standard symplectic reduction, we
get
\beq
\A = \int_{\rm L} \left[ (-\nabla^2 {\tilde \sigma}_{\rm L})\, \delta {\tilde \theta}_{\rm L}
+ {\tilde \Pi}_{\rm L} \, \delta B_{\rm L} \right]
+ \int_{\del {\rm L}}  \E_{0}\, \delta \theta_{0}  
\label{edge8}
\eeq
The corresponding phase space volume element is
\beq
d\mu_{\rm I} = [d{\tilde \sigma} d {\tilde \theta}]_{\rm L}\,
 [d \E_0\, d\theta_0] \, [ d{\tilde \Pi} d B]_{\rm L} \,  \det (-\nabla^2)_{\rm L}
 \label{edge9}
\eeq
where the determinant is understood to be evaluated with Dirichlet conditions
on its eigenfunctions.
The action, the Hamiltonian and the constraints in (\ref{edge4}) are given by
\beqar
\S_{\rm L} &=& \int_{\rm L} \left[ (-\nabla^2 {\tilde \sigma}_{\rm L})\, {\dot {\tilde \theta}}_{\rm L}
+ {\tilde \Pi}_{\rm L} \, {\dot B}_{\rm L} \right]
+ \int_{\del {\rm L}}  \E_{0}\, {\dot \theta}_{0} (x) 
- \int dt \,\H \label{edge10}\\
\H &=& \int  {1\over 2}\ \left[  (\nabla {\tilde \sigma}_{\rm L})^2 
+ ( \nabla{\tilde \Pi}_{\rm L} )^2 + B_{\rm L}^2 \right] + {1\over 2} \int_{\del {\rm L}}  \E_{0}(x)
M_{\rm L}^{-1}(x, y) \,\E_{0} (y)
\nonumber
\eeqar
\beq
\delta(\nabla\cdot E) = (\det(-\nabla^2)_{\rm L})^{-1}\, \delta ({\tilde\sigma}_{\rm L})
\hskip .2in
\delta(\nabla\cdot A)  = (\det(-\nabla^2)_{\rm L})^{-1}\, \delta ({\tilde\theta}_{\rm L})
\label{edge11}
\eeq
Using these results and carrying out integrations, including over
$B_{\rm L}$, we find
\beqar
 Z_{\rm L} &=& \int [d\E_0 d\theta_0] \, [d {\tilde \Pi}_{\rm L}] ~\exp ({i {\tilde S}} ) \nonumber\\
 {\tilde \S}_{\rm L} &=&\int {1\over 2} \left[ {\dot {\tilde\Pi}}_{\rm L}^2 - ( \nabla {\tilde \Pi}_{\rm L} )^2 \right]
 + \int_{\del{\rm L}} \left[ \E_{0}\, {\dot \theta_{0 }} - {1\over 2} \E_{0} \, M_{\rm L}^{-1}\, \E_{0}
 \right]
 \label{edge12}
 \eeqar
We see that the dynamics has been reduced to that of a scalar field
${\tilde \Pi}_{\rm L}$ (which obeys Dirichlet conditions on all
of $\del{\rm L}$) and ``edge modes" with dynamics
given by the second term in ${\tilde\S}_{\rm L}$.
 
We now carry out exactly the same analysis for the full space
$\M$. This will give an expression similar to (\ref{edge12}), but for the whole space, and with no edge modes since we chose Dirichlet conditions
on $\del \M$. However, we can also choose to split any field into 
${\tilde \chi}_{\rm L}$, ${\tilde \chi}_{\rm R}$ and $\chi_0$ with
\beq
\chi = \begin{cases}
 {\tilde \chi}_{\rm L}(x) + \int_{\del {\rm L}} \chi_{0} (y)\,n \cdot \del G_{\rm L} (y,x)\hskip .1in&{\rm in~L~and~on}~\del{\rm L}\\
 {\tilde \chi}_{\rm R}(x) + \int_{\del {\rm R}} \chi_{0} (y)\,n \cdot \del G_{\rm R} (y,x)\hskip .1in&{\rm in~R~and~on}~\del{\rm R}=\del{\rm L}\\
\end{cases}
\label{edge13}
\eeq
With this parametrization of the fields, the action takes the form
\begin{align}
\S_{\rm split} 
&=\int_{\rm L} \left[ (-\nabla^2{\tilde \sigma}_{\rm L} ) \,{\dot{\tilde \theta}}_{\rm L} -{1 \over 2} (\nabla \tilde{\sigma}_{\rm L})^2 \right]+ \int_{\rm R} \left[
(-\nabla^2{\tilde \sigma}_{\rm R} )\, {\dot{\tilde \theta}}_{\rm R} -{1 \over 2} (\nabla \tilde{\sigma}_{\rm R})^2 \right] \nonumber\\
&+ \int   \Pi {\dot B} - {1 \over 2} \left[ (\nabla \Pi )^2 + B^2 \right]
+ \int_{\del{\rm L}} \E_0 \,{\dot \theta}_0  -
{1\over 2} \E_0 (M_{\rm I} + M_{\rm II} )^{-1} \E_0 
\label{edge14}
\end{align}
From this, we can read off the canonical one-form
\beq
\A = \int_{{\rm L}\cup{\rm R}} \Pi\, \delta B + \int_{\rm L} (-\nabla^2 {\tilde \sigma}_{\rm L}) \,\delta {\tilde \theta}_{\rm L}
+\int_{\rm R} (-\nabla^2 {\tilde \sigma}_{\rm R}) \delta {\tilde \theta}_{\rm R}
+ \int_{\del{\rm L}} \E_0\, \delta \theta_0
\label{edge15}
\eeq
The phase volume associated to this is
\beq
d\mu_{\rm split}  = [d{\tilde \sigma} d{\tilde \theta}]_{\rm L} \,[d{\tilde \sigma} d{\tilde \theta}]_{\rm R} \,\det (-\nabla^2)_{\rm L} \, \det (-\nabla^2)_{\rm R}\,
[d \E_0 d\theta_0] \, [d\Pi dB]
\label{edge16}
\eeq
Notice that $d\mu$ involves $\det (-\nabla^2)$ calculated separately for ${\rm L}$ and ${\rm R}$ with Dirichlet conditions.
The determinant $\det(-\nabla^2)$ for the full space appearing
in the path integral (as in (\ref{edge4}))
can also be displayed in the split form using the
BFK gluing formula as\cite{BFK}
\beq
\det (-\nabla^2) = \det (-\nabla^2)_{\rm L} \, \det (-\nabla^2)_{\rm R} \, {\det (M_{\rm L}
+ M_{\rm R})}
\label{edge17}
\eeq
As regards the constraints, we can write them as
\beqar
\int \del_i  f E_i &=& \int_{\rm L} \tilde{f }_{\rm L}(-\nabla^2 {\tilde \sigma}_{\rm L}) + 
 \int_{\rm R} \tilde{f}_{\rm R} (-\nabla^2 {\tilde \sigma}_{\rm R})  + \int
 f_0 \, \E \approx 0\label{edge18}\\
\int \del_i  h A_i &=& \int_{\rm L} \tilde{h}_{\rm L} (-\nabla^2 {\tilde \theta}_{\rm L}) + 
 \int_{\rm R} \tilde{h}_{\rm R} (-\nabla^2 {\tilde \theta}_{\rm R})  + \int
  h_0 \, (M_{\rm L} + M_{\rm R} )\,\theta_0  \approx 0
\nonumber
\eeqar
If we plan to integrate over the full space, the constraints eliminate
$\theta_0$-dependence everywhere, which is equivalent to writing
\beqar
\delta (\nabla \cdot E)\,\delta(\nabla \cdot A) &=& \delta [ -\nabla^2{\tilde \sigma}_{\rm L}]\,\delta [ -\nabla^2{\tilde \sigma}_{\rm R}]\,
~ \delta [-\nabla^2{\tilde \theta}_{\rm L}]\,\delta [ -\nabla^2{\tilde \theta}_{\rm R}]\nonumber\\
&&\times
\delta [ \E_0 ] \delta [(M_{\rm L} + M_{\rm R})  \theta_0 ]\label{edge19}\\
&=&\delta \left[ \cdots \right]\, \det(-\nabla^2)_{\rm L}^{-1} 
\det (-\nabla^2)_{\rm R}^{-1} \det (M_{\rm L} + M_{\rm R} )^{-1}\nonumber
\eeqar
where $\delta \left[ \cdots \right]$ indicates the product of
$\delta$-functions for fields ${\tilde \sigma}_{\rm L}$,
${\tilde \sigma}_{\rm R}$, ${\tilde \theta}_{\rm L}$, ${\tilde \theta}_{\rm R}$,
$\E_0$, $\theta_0$.
Notice again the appearance of determinants separately for ${\rm L}$ and
${\rm R}$, but there is also one factor of 
$\det (M_{\rm L} + M_{\rm R} )^{-1}$.
This arises from the last term in the second of the constraints
(\ref{edge18}).
It is then easy to check that the path integral for the full space is 
reproduced. 

However, if we now consider integrating over fields in
${\rm R}$, the edge modes $\E_0$ and $\theta_0$ on the interface
are physical degrees of freedom from the point of view of
region ${\rm R}$. The test functions $f_0$, $h_0$ in (\ref{edge18}) 
are to be taken to be zero, so we do not have
the corresponding $\delta$-functions or the factor 
$\det (M_{\rm L} + M_{\rm R} )^{-1}$ in (\ref{edge19}).
We then find, after integration over fields in ${\rm R}$,
\beqar
Z&=& {\det (M_{\rm L} + M_{\rm R})} \int  
[d\E d\theta_0] [d\Pi d B]\, e^{i \S}\nonumber\\
\S &=& \int  \left[ \E\, {\dot \theta_0} - {1\over 2}  \E ( M_{\rm L}  + M_{\rm R} )^{-1}\,
\E  \right] + \S_{\Pi, B} 
\label{edge20}
\eeqar
The $\Pi, B$ part, which we have not displayed in split-form or elaborated on,
 behaves as a scalar field and gives what is expected for a 
 scalar field as regards entanglement.
 The new ingredient is that, while the result reduces to the expected one
in region ${\rm L}$, {\it there
is an extra factor $\det (M_{\rm L} + M_{\rm R})$ from the phase volume, i.e., an extra degeneracy factor due to the edge modes.}
This contributes to the entanglement entropy, which is now of the form
 \beq
 S_E =  \log \det (M_{\rm L} + M_{\rm R}) + S_{E\,\Pi} 
\label{edge21}
 \eeq
 where $S_{E\, \Pi}$ is due to the $(\Pi, B)$-sector. The extra contribution 
 from the edge modes:\\
 1. is the so-called contact term, calculated many years ago via the replica trick by Kabat \cite.\\
 2. is the interface term needed for the BFK gluing formula.\\
 This is one of the key effects of the edge modes we wanted to emphasize.
 For more details and extensions to the nonabelian, as well as Maxwell-Chern-Simons theories, see reference [3].
 
 \section{The EIH method for Chern-Simons + Einstein gravity}
 
 We now turn to another effect of edge modes, namely, how they play a role in the context of the Einstein-Infeld-Hoffmann (EIH) method \cite{EIH}. Recall that EIH argued 
 that the Einstein field equations for gravity in the vacuum are sufficient to determine the dynamics of point-particles (i.e., matter dynamics)
 interacting gravitationally.
They defined point-particles as singularities in the gravitational field,
excised small spheres (or tubes when we include time)
around the singularities to keep fields
 well-defined, and imposed the field equations on the resulting configurations.
 This led to a set of equations which are not only conceptually interesting, but actually can be applied in some astrophysical contexts.
 The question we pose here is: How does this work in more general theories of gravity, say, Chern-Simons (CS) gravity or in CS gravity with an Einstein-Hilbert term added \cite{Jiusi-N}? To begin with let us consider the 2+1 dimensional CS action, with connections in the algebra of some Lie group $G$, given as
\beq
S = {k \over 4 \pi} \int \Tr ( A \, dA + {\textstyle{ 2\over 3}} A^3 ) + S_b (A, \psi )
\label{edge22}
\eeq
 The spacetime manifold is taken as $M \times \mathbb{R}$, where
 $M$ has the topology of the disc. $S_b (A, \psi )$ is a boundary action
 (which may depend on some other fields $\psi$) which ensures the full gauge invariance of the action (\ref{edge22}), including transformations
 on the boundary $\del M$.
 
 The bulk equation of motion is $F = 0$, i.e., $A$ is a pure gauge in the bulk.
 Consider now singular classical solutions on the disk $M$ of the form
\beq
A_i = a_i , \hskip .3in da + a^2 = \sum_{s =1}^N  q_s \,\delta^{(2)}(x - x_s),
\hskip .3in
A_0 = a_0 = 0
\label{edge23}
\eeq
 For simplicity, we take all $q_s$ to be in the Cartan subalgebra of $G$
 so that $d a + a^2 = da$. We still have $F =0 $ on ${\tilde M} = M -
 \{ C_s \}$ where $C_s$ denote small disks around the singularities;
 thus $A$ is still a pure gauge on ${\tilde M}$. The general solution to the field
 equation is then a gauge transform of (\ref{edge23}),
 \beq
A_i = g^{-1} a_i \, g + g^{-1} \del_i g , \hskip .3in
A_0 = g^{-1} \del_0 g 
\label{edge24}
\eeq
The evaluation of the action on this configuration gives
\beq
S = S[ a] - {k \over 4 \pi} \sum_s \oint_{\del C_s} \Bigl[\Tr (a \, dg \, g^{-1})\Bigr]
\label{edge25}
\eeq
We see that the dynamics is given in terms of the group elements
$g$ on $\del C_\alpha$; these represent the ``edge modes".
Consider shrinking the size of the disks to almost zero radius, so that
$g$ can be taken to be $g_s = g(\vx_s )\equiv h^{-1}_s$. The $g$-dependent part of the action, after integrating over the spatial boundaries,
 is thus
\beq
S =  {k \over 4 \pi} \sum_s \int dt \Bigl[\Tr (q_s \, h^{-1}_s \del_0 h_s )\Bigr]
\label{edge26}
\eeq 
This is a sum of co-adjoint orbit actions\footnote{another of Bal's favorite topics}. We see that there is no real dynamical evolution, not surprising for the CS theory, but, in the usual manner of quantizing such actions, the states in the quantum theory
will carry unitary representations of $G$, of highest weights specified
by the $q_s$.

With this observation in mind, consider Einstein gravity in 2+1 dimensions
(with a cosmological constant $\sim l^{-2}$) which may be described by the CS action\cite{3dgrav}
\beqar
S &=& -{k \over 4\pi} \left[\int \Tr \left( A dA + {\textstyle{2\over 3}} A^3\right)_L
- \int \Tr \left( A dA + {\textstyle{2\over 3}} A^3\right)_R \right]\nonumber\\
&=&  -{k \over 4\pi \, l} \int d^3x\, \det e~\left[ R - {2 \over l^2} \right]
+ {\rm total~derivative}
\label{edge27}
\eeqar
where the gauge connections are given in terms of the spin connection
$\omega^{ab}$ and the frame field (dreibein in this case)  $e^a$ by
\beqar
A_L &=& (-i M_a) \, A^a_L = (-i M_a) \left( -\half \eta^{ak} \e_{kbc} \omega^{bc}  + {e^a\over l} \right)\nonumber\\
A_R &=& (-i N_a) \, A^a_R = (-i N_a) \left( -\half \eta^{ak} \e_{kbc} \omega^{bc}  - {e^a\over l} \right)
\label{edge28}
\eeqar
 Here $M_a$, $N_a$ are the generators of two independent
 $SO(2, 1)$ Lie algebras, with parity exchanging them.
 Newton's constant for this problem can be identified as
 $G= l/(4 k)$.
 
Since there is a nonzero cosmological constant, it is
clear from (\ref{edge27}) that the solution for the vacuum state, i.e., the solution of the bulk equation of motion, is the anti-de Sitter (AdS) space in 2+1 dimensions.
 As with (\ref{edge22}), we can now add a boundary term (on $\del M$) to 
 obtain full gauge invariance for (\ref{edge27}) and then introduce
 ``point-particles" via
 ans\"atze of the form (\ref{edge23}), (\ref{edge24}). The action on these
 configurations becomes
 \beq
S = - {k\over 4\pi} \sum_s \int dt~\left[ \Tr (q_s \,  h^{-1}_s\,{\del_0 h}_s)_L
- \Tr (q_s \,  h^{-1}_s\,{\del_0 h}_s )_R\right]
\label{edge29}
\eeq
We get two sets of co-adjoint orbit actions, leading to unitary representations
of $SO(2,1) \times SO(2,1)$ upon quantization.
The full isometry group for the AdS space
is $SO(2,1) \times SO(2,1)$, with the diagonal $SO(2,1)$ as the Lorentz group, while the coset directions correspond to translations.
Thus a point-particle in AdS space must be defined in the quantum theory
as a unitary irreducible representation of $SO(2,1) \times SO(2,1)$,
and this is exactly what is obtained, realizing the EIH strategy.
The mass and spin, read off from the identification of 
translations and Lorentz transformations, are
\beq
m = {q_R + q_L \over 32 \pi G},\hskip .2in
s = {l \over 16 G} (q_L - q_R)
\label{edge30}
\eeq

In 2+1 dimensional gravity, curvatures are localized at the positions of the particles, so that the result (\ref{edge29}) with no interactions between particles is indeed what we expect. So it is interesting to consider a similar analysis in 4+1 dimensions, where the Einstein-Hilbert action is distinct from a combination of CS terms \cite{zanelli} and can be included as an additional term
in the action. We can construct an $SO(4,2)$ algebra using $4\times 4$
Dirac $\gamma$-matrices, $\gamma_a$,
$\Sigma_{ab} = (i /4) [ \gamma_a, \gamma_b ]$, $ a, b =
0, 1, 2, 3, 5$.  We then define
the connections
\beq
A_L = - {i\over 2} ( \omega^{ab} \Sigma_{ab} + e^a \gamma_a ),
\hskip .2in
A_R =  - {i\over 2} ( \omega^{ab} \Sigma_{ab} - e^a \gamma_a )
\label{edge31}
\eeq
As for the action, we will consider the parity-invariant combination
\beqar
S&=& CS (A_L) - CS(A_R) + S_b ( A, \psi) + S_{\rm EH}\nonumber\\
CS(A)&=& - {i k \over 24 \pi^2} \int
\Tr \left[ A dA dA + {3 \over 2} A^3 dA + {3 \over 5} A^5 \right]
\label{edge32}
\eeqar
$S_b (A, \psi)$ is, as before, included to cancel any boundary terms from
gauge transformations, and $S_{\rm EH}$ is the Einstein-Hilbert action 
with a cosmological constant $\sim l^{-2}$. In terms of the curvature 
$R^{ab} = d \omega^{ab} + (\omega \omega)^{ab}$ and 
torsion $T^a = d e^a + \omega^{ab} e^b$, we find
\beq
F_{L, R} = (- i \Sigma_{ab} /2) (R^{ab} + e^a e^b ) \pm
(-i \gamma_a ) T^a
\label{edge33}
\eeq
The bulk equation of motion for the $CS$-part in (\ref{edge32}) is thus satisfied by $R^{ab} = - e^a e^b$ and $T^a = 0$. With the scaling,
$e^a \rightarrow \sqrt{12}\, e^a/l$, we see that this is AdS spacetime.
For simplicity, we take $S_{\rm EH}$ to have the same value for the cosmological constant so that the 
term from $S_{\rm EH}$ in the bulk equation of motion is also zero for the same AdS spacetime.

We now introduce solutions with point-like singularities on the spatial manifold. These will be taken as the point-like limit of 4d-instantons 
in $SO(4) \in SO(4,2)$. We excise small balls around each point,
and on $M - \{ C_s \}$, we take the solution to be of the
form
\beqar
a &=& t_1 \, U^{-1} d U, \hskip .2in
t_1= \left( \begin{matrix} 1&0\\
0&0\\ \end{matrix}\right), \hskip .1in {\rm or}\hskip .1in
\left( \begin{matrix} 0&0\\
0&1\\ \end{matrix}\right)\nonumber\\
U &=&  \phi^0 + i \sigma_i \phi^i, \hskip .2in
\phi^0 \phi^0 +  \phi^i \phi^i = 1
\label{edge34}
\eeqar
$t_1$ is a projector to either of the two $SU(2)$'s in
$SO(4) \sim SU(2) \times SU(2)$.
The instanton number is the integral of $(1/12\pi^2)\epsilon_{\mu\nu\alpha\beta}\phi^\mu d \phi^\nu d \phi^\alpha d \phi^\beta$. We then consider the more general configurations with the same singularity structure:
\beq
A_L = g^{-1} ( a_L + \delta a_L) g + g^{-1}  dg , \hskip .2in
A_R = g^{1} (a_R + \delta a_R) d g + g^{-1} dg 
\label{edge35}
\eeq
$a_L$ and $a_R$ are terms of the form (\ref{edge34}), and are
related by parity. $\delta a_L$ and $\delta a_R$ correspond to 
perturbations of the metric as $g_{\mu\nu} \rightarrow g_{\mu\nu} + h_{\mu\nu}$, i.e.,
\beq
\delta a_{L,R} = (-i/2) \left[ \pm \gamma_a\, e^{-1 \nu a} h_{\mu\nu}
- \Sigma_{ab} \, e^{-1 \alpha a} e^{-1 \beta b} \nabla_\alpha h_{\beta \mu}
\right] dx^\mu
\label{edge36}
\eeq
Also $g$ is an element of $SO(4,2)$ parametrized as
\beq
g = S^{-1} \Lambda V, \hskip .2in
V = {1\over \sqrt{z}}  \left( \begin{matrix}
z ~& i X\\
0 ~& 1\\ \end{matrix}\right), \hskip .1in S = {1\over \sqrt{2}}
\left( \begin{matrix}
1&1\\ 1&-1\\ \end{matrix} \right)
\label{edge37}
\eeq
Here $X= x^0 - i \sigma_i x^i$, and $z = x^5$ is the radial coordinate of AdS
and $\Lambda \in SO(4,1)$ denotes a Lorentz transformation. Evaluating the action (\ref{edge32}) on the configurations (\ref{edge35}), we find
\beq
S = {k \over 2} \sum_s \int (Q^1_s - Q^2_s ) \eta_{ab}
(\Lambda_s)^a_0 \left[ e^b_\mu + 
e^{-1 \nu b} h_{\mu\nu} \right] dx^\mu
+ {1\over 2}\int  h_{\mu\nu} \mathbb{L}^{\mu\nu \alpha \beta} h_{\alpha \beta}
\label{edge38}
\eeq
where $Q^1_s$ and $Q^2_s$ are the instanton numbers for the two $SU(2)$'s in $SO(4)$ and $\mathbb{L}^{\mu\nu \alpha \beta} $ is the Lichnerowicz operator
(i.e., the quadratic fluctuation operator) for $S_{\rm EH}$. Without the
$h_{\mu\nu}$-terms, this is a sum of co-adjoint orbit actions. The mass may be
identified as $m_s = -\half k (Q^1_s - Q^2_s)$. (Spin can be included
by considering different orientations of the $SU(2)$'s in $SO(4)$ for different instantons.)
Further, if
we write $\Lambda^a_0 = e^a_\nu {dx^\nu\over ds}$, which is consistent
with its properties, the first term in $S$ becomes $- m \int ds$, corresponding to free particle motion. Solving for $h_{\mu\nu}$ from its equation of motion {\it \`a la} (\ref{edge38}), and using it back again in (\ref{edge38}), we get
interactions between the particles. In the nonrelativistic limit, the result is
\beq
S = -\sum_{s} \int m _s \int ds_s + {1\over 6} \sum_{s\neq s'}\int dx^0_s
{m_s m_{s'} \over 4 \pi^2 r^2_{s s'} } , \hskip .1in x_{s'}^0 =
x_s^0 + \vert \vx_s - \vx_{s'}\vert
\label{edge39}
\eeq
showing, correctly, the 4d Coulomb-like potential, if we neglect retardation effects.
In principle, one can include higher corrections systematically, but this suffices to prove our main point: The edge modes of the CS+Einstein gravity
allow us to realize the Einstein-Infeld-Hoffmann method of defining 
point-particles as singularities of the gravitational field and then obtaining their equations of motion from the field equations.

This work was supported in part by U.S. National Science Foundation research grants PHY-2112729 and PHY-1820271.

%%%%%%%%%%%%%%%%%%%%%%%%%%%%%%%%%%%%%%%%%%%%%%%
%%%%%%%%%%%%%%%%%%%%%%%%%%%%%%%%%%%%%%%%%%%%%%%
%%%%%%%%%%%%%%%%%%%%%%%%%%%%%%%%%%%%%%%%%%%%%%%

%%%%%%%%%%%%%%%%%%%%%%%%%%%%%%%%%%%%%%%%%%%%%%%
%%%%%%%%%%%%%%%%%%%%%%%%%%%%%%%%%%%%%%%%%%%%%%%
%%%%%%%%%%%%%%%%%%%%%%%%%%%%%%%%%%%%%%%%%%%%%%%
%%%%%%%%%%%%%%%%%%%%%%%%%%%%%%%%%%%%%%%%%%%%%%%

\begin{thebibliography}{99}
\bibitem{gen} Entanglement is a topic with an enormous literature by now, it is impossible to make even a vaguely comprehensive list in the short space available. 
The papers cited below can be used to trace most of the relevant articles.
\end{thebibliography}
\end{document}